%% file: main.tex
\documentclass[utf8]{FrontiersinHarvard} 
\usepackage{url,hyperref,lineno,microtype,subcaption, graphicx}
\usepackage[onehalfspacing]{setspace}

\def\firstAuthorLast{Rack {et~al.}} 
\def\Authors{Christian Rack\,$^{1,*}$, Tamara Fernando\,$^{1}$, Murat Yalcin\,$^{1}$, Andreas Hotho\,$^{2}$ and Marc Erich Latoschik\,$^{1}$}

\def\Address{$^{1}$Human-Computer Interaction (HCI) Group, Informatik, University of Würzburg, Würzburg, Germany\\
$^{2}$Data Science Chair, Informatik, University of Würzburg, Würzburg, Germany}

\def\corrAuthor{Christian Rack}

\def\corrEmail{christian.rack@uni-wuerzburg.de}
\usepackage{tabu}
\usepackage{booktabs}
\usepackage{pgf}
\usepackage{multirow}
\usepackage{amsmath}
\usepackage{tabularx}
\usepackage{cleveref}
\usepackage{siunitx}
\usepackage{orcidlink}
\usepackage{float}

\usepackage[commentmarkup=footnote,final]{changes}
\definechangesauthor[name=Marc, color=orange]{ml}
\definechangesauthor[name=Andreas, color=red]{ah}
\definechangesauthor[name=Chris, color=violet]{cs}

\newcommand{\acc}[2]{$Acc^{\text{enr}: #1}_{\text{use}: #2}$}

\begin{document}

\onecolumn
\firstpage{1}

\title{Who Is Alyx? A new Behavioral Biometric Dataset for User Identification in XR}

\author[\firstAuthorLast ]{\Authors} 
\address{} 
\correspondance{} 

\extraAuth{}

\maketitle

\begin{abstract}
\section{}
This article presents a new dataset containing motion and physiological data of users playing the game `Half-Life: Alyx'.
The dataset specifically targets behavioral and biometric identification of XR users.
It includes motion and eye-tracking data captured by a HTC Vive Pro of 71 users playing the game on two separate days for 45 minutes.
Additionally, we collected physiological data from 31 of these users.
We provide benchmark performances for the task of motion-based identification of XR users with two prominent state-of-the-art deep learning architectures (GRU and CNN).
After training on the first session of each user, the best model can identify the 71 users in the second session with a mean accuracy of 95\% within 2 minutes.
The dataset is freely available under \url{https://github.com/cschell/who-is-alyx}
\end{abstract}


\input{sections/01_introduction}
\input{sections/02_related_work}

\input{sections/03_methods}
\input{sections/04_results}
\input{sections/05_discussion}
\input{sections/99_figures_and_tables}

\bibliographystyle{Frontiers-Harvard}
\bibliography{references,references_zotero_chris}

\end{document}

%% file: sections/01_introduction.tex
\section{Introduction}

The field of Virtual Reality, Augmented Reality and Mixed Reality (VR, AR and MR; extended reality (XR), in short) matured considerably in the last decade and enables new ways for corporations, institutions and individuals to meet, support and socialize without the need for users to be physically present.
Consequently, XR architects are increasingly faced with the challenge of developing solutions that promote authenticity and trust among users, while meeting the high usability requirements of XR applications at the same time.
For example, typical password-based verification mechanisms require text input, which is problematic to provide in XR contexts as it comes with notable usability problems~\citep{Stephenson2022,Dube2019,Knierim2018,Kern2021}.
This renders typing-based security mechanisms cumbersome and sub-par.

Recent approaches explore behavioral biometrics as an interesting alternative solution for identification and verification.
In addition, such approaches open-up novel features for social XR.
For example, they allow continuous monitoring of user movements to provide important cues about avatars' identities, i.e., who is controlling the digital alter egos, which becomes increasingly important when avatars are replicating the appearance of real people \citep{Achenbach2017,Wenninger2020}.
Such approaches do not necessarily need extra hardware, but can work with the data that is already provided freely by the XR system.
Of particular interest in this context is the tracking data that captures the movements of XR users, as this is a constant and readily available data stream that is required for any XR system in order to provide an immersive experience.
Analogous to the handwriting of a person, movements in general seem to have very individual characteristics: previous works have demonstrated that identifying information can be extracted from users' movements, so using tracking data provides the foundation for novel approaches to identify and verify XR users.

Following \cite{biometrics_intro}, systems of user recognition can generally be divided into two categories: identification and verification. 
Identification systems involve the use of biometric inputs from a user to query a database containing the biometric templates of all known users, effectively seeking to answer the question, 'Who is this?'. 
Verification systems, on the other hand, compare the biometric input against the templates associated with a specific identity claimed by the user.
In this context, `verification' is synonymous with `authentication', akin to password-based verification where a password is verified against a specific user's known credentials, not against all users in the database.
This distinction, although often blurred in the literature, is crucial for this article, as these two concepts embody different methodologies, each with its own considerations and applications.
This being said, verification and identification approaches are still related on a technical level and can share the same underlying core mechanics (e.g., data preprocessing, machine learning architecture, etc.).

Recent efforts to leverage user motions for identification or verification purposes rely mainly on machine learning models.
While these models yield promising results, the works proposing them become susceptible to the ongoing general discussion about a growing reproducibility crisis concerning machine learning methods in research~\citep{Kapoor2022}.
One requirement is that researchers must choose the correct methodology and the right evaluation methods to avoid overoptimistic conclusions about their work.
Another requirement is reproducibility, as otherwise understanding proposed solutions becomes difficult, errors cannot be found, and new solutions cannot be compared to previous ones.
In general, there is a trend that non-replicable publications are cited more often than replicable ones \citep{Serra-Garcia2021}, which adds to the overall problem.
An important component for researchers to make their work reproducible is to use common, publicly available datasets~\citep{Pineau2020}.

Unfortunately, there are very few \textit{public} datasets for research concerning XR identification methods:
to our knowledge, there are only three datasets featuring tracking data of XR users  \citep{Rack2022ComparisonSequences,Miller2022CombiningBiometrics,Liebers2021}. 
These datasets are relatively small, or just single-session, which makes them an inadequate choice for the development and the evaluation of larger (deep) machine learning models.
This is a severe problem, since the lack of public datasets makes it impossible for researchers 
    1) to participate in this domain of research if they do not have the time, resources, or access to XR users to create their own dataset,
    2) to compare their new methods to current ones, and
    3) to reproduce previous solutions.
This is in stark contrast to other fields of research, where large public datasets have facilitated the development of impressive systems like BERT (natural language processing)~\citep{Devlin2019}, YOLO (object detection) \citep{Wang2021} or the recent DALL-E2 (text to image)~\citep{Ramesh2022} and ChatGPT~\citep{gpt4}.

Consequently, we see the lack of public datasets as a serious barrier for mature and practical solutions in this area of research.
With our work, we address this issue and publish a new dataset designed for XR user recognition research.
The dataset includes the tracking data of 71 VR users who played the game `Half-Life: Alyx' with a HTC Vive Pro for 45 minutes on two separate days.
In addition to recording motion data and eye-tracking from the VR setup, we also collected physiological data for a subset of 31 users, which may be interesting for future work as well.
Since the motivation for this dataset is to serve future research for XR user identification and verification, we benchmark two state-of-the-art machine learning architectures that have been shown to find identifying characteristics in user motions.

Recognizing the lack of public datasets as a serious barrier for mature and practical solutions in this area of research, our work provides a new, specially designed dataset for XR user identification research.
The dataset includes tracking data from 71 VR users who played the game `Half-Life: Alyx' on an HTC Vive Pro for 45 minutes on two different days.
Alongside the motion and eye-tracking data from the VR setup, we also collected physiological data from a subset of 31 users.
While our dataset can potentially be used for both identification and verification research, the focus of this paper is on user identification.
To demonstrate the practical value of this dataset, we apply two state-of-the-art machine learning architectures, showing their capacity to identify unique characteristics in user motion data.

Altogether, our contributions are as follows:

\begin{enumerate}
\item A new public dataset of 71 VR users, with two sessions per user of 45 minutes each.
\item Validation of the dataset for the identification task with two state-of-the-art deep learning architectures (CNN and GRU), providing benchmark results for future work.
\item We publish all data and code: \url{https://github.com/cschell/who-is-alyx}
\end{enumerate}

%% file: sections/02_related_work.tex
\section{Related Work}

Our dataset is different to most used by related work, as it is not only freely accessible and relatively large, but also provides a unique setting with VR users performing a wide range of non-specific actions.
In the following we discuss the landscape of datasets used by previous works and explain how the ``Who Is Alyx?'' dataset compares.

\subsection{Specific vs. Non-Specific Actions}
\label{sec:specific-vs-non-specific}

In this article, we use the terms ``specific action'' and ``non-specific action" to describe the contextualization of input sequences for identification models.
\textit{Specific actions} refer to highly contextualized input sequences where the user action is well known and consistent.
\textit{Non-specific actions} on the other hand result in less contextualized input sequences that include more arbitrary user motions.
We categorize previous work based on the degree of contextualization of input sequences.

\textit{Highly contextualized} sequences, where user actions are well-defined, are explored by a wide range of works.
Here, users are tasked to throw or re-locate virtual (bowling) balls \citep{Miller2021, Ajit2019, Kupin2019, Liebers2021, Olade2020}, to walk a few steps \citep{Pfeuffer2019, Shen2019, Mustafa2018}, to type on a virtual keyboard \citep{Mathis2020, Bhalla2021} or to nod to music \citep{Li2016}.
In each of these scenarios, the identification model receives short input sequences that include a well defined user action, so most of the variance in the input data should be attributable to the user's personal profile.
This allows identification models to specialize on user-specific characteristics within particular motion trajectories.
Such scenarios are interesting for authentication use-cases, where the user is prompted to do a specific action and the system can perform the verification based on the same motions each time.

\textit{Medium contextualized} input sequences originate from scenarios that allow more freedom for users to move and interact.
For example, \cite{Miller2020a} identified users watching 360° VR videos or answering VR questionnaires for several minutes.
Following this work, \cite{Moore2021PersonalSessions} explored an e-learning scenario where VR users learned to troubleshoot medical robots in several stages.
And recently, \cite{Rack2022ComparisonSequences} compared different classification architectures using a dataset of participants talking to each other for longer periods.
In each of these cases, the general context of the scenario still limits possible user actions (e.g., to `listening' and `talking'), but the exact action, its starting, and its ending point become uncertain.
Consequently, the identification task becomes more challenging as models have to deal with more variance not related to the user's identity.
Also, it becomes necessary to observe the user for longer periods (i.e., minutes) to make reliable identifications.
However, models that can detect user-specific characteristics within such input sequences promise to be applicable to a wider range of scenarios where individual user actions are not clearly specified anymore.
This becomes interesting for use cases where XR users cannot be prompted to perform a specific identifying motion and instead need to be monitored passively.

The ``Who is Alyx?'' dataset is \textit{marginally contextualized}.
During recording, the user plays the game for 45 minutes and executes a wide range of actions along the way.
Even though there are several actions that occur repeatedly, like using tools, reloading or using the teleportation mechanism, we consider them non-specific:
input sequences for the model could be taken from anywhere in the user recording and may include all, some, or none of these actions.
Hence, identification models have to deal with a lot of variance and have to learn user-specific patterns within sequences from a wide range of more or less arbitrary user motions.
This makes marginally contextualized datasets like ours are an obvious choice to develop identification models that can be applied to a wide range of XR scenarios and use cases.

\subsection{Existing Datasets}

Besides providing a wide range of user actions, the ``Who is Alyx?'' dataset is also the largest freely available dataset featuring XR user motion data.
Although there is already a sizeable body of literature investigating behavior-based identification, to date there are to our knowledge only three \textit{publicly} available datasets viable for the development of methods to identify XR users by their motions.
\Cref{tab:existing_datasets} lists these datasets and compares them with ``Who is Alyx?''.

\cite{Liebers2021} published a dataset with 16 VR users, each performing the two specific actions ``throwing a virtual bowling ball'' and ``shooting with bow and arrow''.
In each of two sessions, users had to complete 12 repetitions of both actions using an Oculus Quest.
Each repetition took between 5 and 10 seconds on average, resulting in a total recording time of 4 minutes per user per action.

\cite{Rack2022ComparisonSequences} use the public ``Talking With Hands'' dataset of \cite{Lee2019} to identify 34 users. 
The dataset provides full-body motion tracking recordings from participants talking to each other about previously watched movies.
After filtering and cleaning the data, the authors used about 60 minutes per user to train and evaluate their models.

\cite{Miller2022CombiningBiometrics} recently published a combination of two datasets, featuring 41 and 16 users.
These datasets are similar to those of \cite{Liebers2021}:
they contain recordings of users performing a specific action (i.e., throwing a ball) on two separate days.
Additionally, the users repeated the action with three different VR devices, which allows evaluations regarding cross-device generalizability.
The datasets contain for each user 10 repetitions of ball-throwing per session and device, each lasting for exactly 3 seconds, resulting in a total of 1 minute of footage per user and device and session.

Each of these datasets has considerable drawbacks for developing and testing new deep learning approaches.
The datasets of  and \cite{Miller2022CombiningBiometrics} provide only a small amount of data per user and session.
This is not just a problem because deep learning models are prone to learn wrong characteristics from individual repetitions, but also because there is too little test data to draw meaningful conclusions about the model's ability to generalize well to unseen data.
For example, we replicated the work of \cite{Liebers2021} and followed their proposed methods to train an LSTM model.
Here, the test results changed considerably by $\pm 8$ percentage points just by repeating training with differently seeded weights, suggesting that either the training data is too small for robust training or the evaluation data is insufficient to reflect how well the model generalizes.
In contrast, the dataset used by \cite{Rack2022ComparisonSequences} provides much more footage per user, but is only single-session, so it cannot be tested how well a model would recognize the same person on a different day.

Against this backdrop, we created the ``Who is Alyx?'' dataset with the purpose of not only being publicly available, but to improve upon existing free datasets by providing much more data and two sessions for each user.
In this article we demonstrate that current state of the art identification models can be robustly trained on session one and generalize to session two.
This provides future work not only with a large dataset, but also with benchmark scores to compare new solutions with.

Note that just before the submission of this work Vivek et al. published their BOXRR dataset with motion data from over 100,000 online players of the VR games ``Beatsaber'' and ``Tilt Brush'' \citep{BOXRR-23}.
Despite this exciting development, the ``Who is Alyx" dataset continues to be a highly valuable resource for researchers investigating XR user behavior and biometric identification. While our dataset comprises a smaller cohort of 71 users, it still stands out due to several distinctive attributes.
Firstly, the data collection process was conducted in a controlled environment, ensuring precise hardware setup and confirming each participant's active involvement in the XR experience of Half-Life: Alyx on two separate days.
Notably, the ``Who is Alyx'' dataset extends beyond mere motion tracking, encompassing not only screen recordings, but additional biometric data such as eye-tracking and physiological signals.
This comprehensive approach offers a holistic perspective on users' physiological responses during XR interactions.
Lastly, the two datasets vary in levels of contextualization in the actions performed by users 
between highly- to medium contextualized (BOXRR) and marginally contextualized (ours) actions, diversifying the data available to the research community.
As a result, despite the lower overall user count, the ``Who is Alyx'' dataset contributes essential and complementary insights to the publicly available XR datasets for researchers investigating XR user identification, biometrics, and behavioral analysis.

%% file: sections/03_methods.tex
\section{Dataset}

In the following we describe the creation process of the dataset and provide details about the VR game.

\subsection{The Game}

We selected the action game ``Half-Life: Alyx"~\citep{ValveCorporation2020}, because it satisfies two of our main objectives for our dataset.
First, the game was specifically designed for virtual reality and requires a wide variety of interactions from the players.
Second, `Half-Life: Alyx' is very entertaining, which made it easier to recruit enough participants that were motivated not only to keep actively engaged throughout one session but also to come back for the second one.
In fact, the drop out rate was fairly low with just 5 (out of 76) participants attending only one session.

In the game the player assumes the role of Alyx Vance, who has to rescue her father.
Each participant starts the first session in ``Chapter 1'', which introduces the player to the core mechanics of the game: navigation, interacting with the environment, and combat.
For navigation, the player can either walk around within the limits of the VR setup or use the trackpad of the left controller to teleport to another location nearby (\Cref{fig:teleportation}).
Players can grab virtual objects with either hand using the trigger buttons on each controller.
Additionally, the game also provides a mechanism to quickly retrieve distant objects using the character's `gravity gloves':
the player can point at a distant object, press the trigger button (\Cref{fig:gravity_grab_1}), and perform a flicking motion with the hand, which repels the object towards the player (\Cref{fig:gravity_grab_2}).
The player then has to snatch the object out of the air at the right moment using the trigger button again (\Cref{fig:gravity_grab_3}).
For combat situations the player can select a weapon for the right hand and then shoot with the trigger button (\Cref{fig:weapon}).
To reload a weapon, the player has to press the right controller's grip button to eject the empty mag, grab behind the shoulders with the left hand to retrieve a new mag from the virtual backpack, insert it, and pull back the rag (\Cref{fig:weapon_reload}).

After learning these basics, the player ventures on to find Alyx' father.
For this, the game leads through a linear level design and confronts the player with different types of obstacles, such as difficult terrain, bared throughways, or blockading enemies.
To solve these challenges, the player has to find the right way, figure out opening mechanisms or engage in combat.
There are several mechanisms required to unlock doors or activate machines that reoccur 2 to 4 times each per session:
on some occasions, players have to use a hand-scanner to find and fix circuits hidden inside walls (\Cref{fig:circuit}).
To activate upgrade machines, players have to solve 3D puzzles (\Cref{fig:puzzle}).
These machines are not mandatory to progress in the game and can be hidden within the level, so the player can ignore them.
While the nature of these mechanisms remains the same during each encounter, they become increasingly challenging and require a fair degree of trial and error every time.
Additionally, there are a several unique challenges that can require a wide range of actions from the player (like \Cref{fig:unique_puzzle}).

\subsection{Ethical Considerations}
The data acquisition study was approved by the institution's responsible ethics committee.
The participants were recruited through the participant recruitment system of our faculty and gave their full consent to publish the collected and anonymized data.
Every participant was fully informed about the intents and purposes and the procedure of the data acquisition.

\subsection{Setup}

The participants were instructed to play the game with an HTC Vive Pro (with 3 lighthouses) within an $10m^2$ area in our lab.
An instructor was present all the time and monitored the participant in the real and virtual world, ensuring that the game was played continuously.
All participants start their first session in ``Chapter 1'' and are then free to play the game without any further instructions.
After about 45 minutes, we end the session regardless of where the player was in the game at that time.
For the second session, we repeat the same procedure, but change the starting point for each player to ``Chapter 3''.

For the first 41 participants we only recorded the tracking data from the head mounted display (HMD) and controllers, along with the eye-tracking data.
We noticed an interest within our research group to also collect physiological data, as this makes the dataset viable for other research areas as well.
Therefore, we started to have the following 35 participants wear an Empatica E4 wristband and a Polar H10 chest strap.
These wearable devices collect physiological data, are easy to setup, and do not restrict the players in their motions – essential requirements as the dataset continues to be intended primarily for research questions around the identification of XR users by their motions.

\subsection{Recording}
\label{sec:recording}

\Cref{tab:recording-overview} provides an overview of devices, collected data streams, sampling rates, and the number of recordings per device, as well as the captured physiological features.
The sampling rates vary between each data stream, depending on the hard- and software we used for recording.
We recorded the tracking data of the first 37 players with an average frequency of 15 frames per second.
After a revision of our recording script we increased the sampling rate to 90 frames per second and added tracking of the physiological data for the following 34 players.
Since recording the movement data was the main objective, we did not interrupt the session if any of the other data streams failed for any reason, which is why there are some missing eye-tracking and physiological recordings.
In \Cref{tab:recording-overview}, we count recordings that provide at least 15 minutes of footage and document any recording faults in the readme file of the data repository.

To retrieve and record the tracking data from the HTC Vive Pro, we use a Python script and the library ``openvr''~\citep{ChristopherBruns2022}.
The tracking data includes the positions (x,y,z, in centimeters) and rotations (quaternions: x,y,z,w) for the HMD, left and right controller.
The coordinate system is set to Unity defaults, so the `y'-axis points upwards.
We use HTC's ``SRanipal'' to retrieve the eye-tracking data and Unity for calibration and recording.


\subsection{Demographics}

At the beginning of the first session we collect demographic data and body properties.
Altogether there were 36 females and 40 males, all between 21 and 33 years old.
\Cref{fig:demographics} summarizes details about age, body height and VR experience.
We make all of the collected information available in a summary CSV file in the data repository.

\subsection{Insights}

In the following, we focus on the tracking data to stay within the scope of this work and leave the analysis of the eye-tracking and physiological data to future work.
Altogether, we have recorded 71 participants over the course of two sessions, 5 participants attended just one session, so in sum the dataset includes tracking data of 76 different individuals.
Each session is on average 46 minutes long, with the three shortest sessions lasting between 32 and 40 minutes.

While screening the dataset, we tried to gain basic insights into how far the participants showed the same or differing behaviour during playing.
For basic insights, we plot the travel path of the HMDs on the horizontal axes in \Cref{fig:player_travel_path}, which reveals different play styles.
Some players tend to stand relatively still without moving much, while others like to walk around or do extensive head movements throughout one session.
Interestingly, there was no strong correlation between meters traveled and teleportations made, as indicated by a Pearson correlation coefficient of only $0.25$.
\Cref{fig:navigation_shift_swarmplot} compares the navigation behavior of the players in both sessions and shows that the players used the teleportation mechanism more frequently in the second session, indicating a learning effect.
Also, we noticed that male participants generally used the navigation mechanisms more extensively: males moved 6.9 meters and teleported 11 times per minute, while females only moved 5.3 meters and teleported 8.5 times per minute on average.

\section{User Identification Deep Learning Benchmark}
We apply two state-of-the-art deep learning models on the dataset to solve the identification task of XR users by their motions.
For this, we adopt the methodology of previous works \citep{Rack2022ComparisonSequences, Mathis2020} as foundation for an experimental setup that evaluates the generalizability of our models in a multi-session setting.
With the reported results and subsequent analyses of our model, the results can serve as a baseline for future work that use our dataset as benchmark.

\subsection{Methodology}

\subsubsection{Architecture}
For the benchmark, we use a recurrent neural network (RNN), as proposed by \cite{Liebers2021} and \cite{Rack2022ComparisonSequences}, and a convolutional neural network (CNN), as proposed by \cite{Mathis2020} and \cite{Miller2021}.
Both architectures are able to work directly with sequential data and do not require dimensionality reduction along the time-axis during data preprocessing, as would be the case with multilayer perceptrons, random forests, or the like.

Our choice of the specific RNN architecture is guided by \cite{Rack2022ComparisonSequences}, who compared different types of RNNs.
We select the GRU architecture, as it worked equally well as the more commonly used LSTMs, but is a bit faster to train due to a slightly smaller architecture.
For the CNN we select an architecture similar to the `FCNN' proposed by \cite{Mathis2020}.
Additionally, we append a dropout unit to each layer, as we found that this helps to counteract overfitting~\citep{dropout}.
We do not copy the hyperparameter settings used by \cite{Rack2022ComparisonSequences} or \cite{Mathis2020}, but instead perform a hyperparameter search to evaluate a range of different settings (see \Cref{sec:hyperparameter_search}) and report the final configurations for both architectures in \Cref{tab:configurations}.

\subsubsection{Data Encoding}
We follow the methods investigated by \cite{Rack2022ComparisonSequences} for encoding tracking data.
As the authors demonstrate, it is necessary to encode the raw tracking data to remove information that does not contain straightforward user characteristics.
In our work, we compare the following three different data encodings.

Raw tracking data returned by the VR setup is referenced to some origin relative to the VR scene, which Rack et al. refer to as scene-relative (SR) encoded data.
SR data encode a lot of information not directly tied to the identity of the user, such as user position or orientation, which makes it difficult for machine learning models to learn actual user-specific characteristics~\citep{Rack2022ComparisonSequences}.
To remove any information about the exact whereabouts of the user, we transform the coordinate system from SR to \emph{body-relative} (BR):
we select the HMD as the frame of reference and recompute positions and orientations of the hand controllers accordingly.
Since the HMD is always the origin in this coordinate system, its positional features and rotation around the up axis become obsolete and are removed.
This only leaves one quaternion encoding the head's rotation around the horizontal axes.
Subsequently, there are 18 features for each frame: (pos-x, pos-y, pos-z , rot-x, rot-y, rot-z, rot-w) $\times$  (controller-left, controller-right) + (rot-x, rot-y, rot-z, rot-w) of the HMD, all given with respect to the HMD as frame of reference.

Following this, \cite{Rack2022ComparisonSequences} found that computing the velocities for each feature from BR data yields even better results, since this removes more noise from the data that otherwise confuses the models.
This \emph{body-relative velocity} (BRV) data is computed differently for positions and orientations: for positions, we compute the difference between frames for each feature individually.
For rotations, we rotate each quaternion by the corresponding inverted rotation of the previous frame, which yields the delta rotation.
This procedure does not change the number of features, so BRV data also consist of 18 features.

Since the BRV encoding produced the best results for Rack et al., we take this train of thought one step further and evaluate the \emph{body-relative acceleration} (BRA) data encoding.
For the conversion from BRV to BRA we apply the same steps as previously performed for the conversion from BR to BRV.
Like BRV, BRA data consists of 18 features.

\subsubsection{Identification Procedure}
\label{sec:identification}
We train the GRU and CNN to map input motion sequences to one of the 71 users.
For this we use 300 consecutive frames (about 20 seconds) of tracking data of one recording (i.e., one user) as input for the model.
Since recording lengths during testing consist of much more than 300 frames, we employ a majority voting scheme to produce a prediction for the whole sequence:
we retrieve individual subsequences from the whole sequence by moving a sliding window frame by frame, apply the model on each subsequence and in the end predict the user who was most commonly inferred.

\subsubsection{Implementation and Training}

We use the machine learning framework PyTorch Lightning for implementing both network architectures.
Both models are implemented and trained as classification networks with a categorical cross-entropy loss function and the Adam optimizer. 
We train each model for a minimum of 30 epochs until the monitored minimum accuracy on the validation set deteriorates or stagnates.
We save a snapshot of each model at its validation highpoint for later evaluation, since performance can already start declining towards the end of the 30 epochs. 
We use PyTorch Lightning \citep{Falcon_PyTorch_Lightning_2019} for implementation. 
Each job runs with 8 CPU units, one GPU (either NVIDIA GTX 1080Ti, RTX 2070 Ti or RTX 2080Ti), and 20 Gb RAM.

\subsection{Experimental Setup}

In our setup we consider a scenario where our system observes VR users playing the game Half-Life: Alyx for some time on one day, and then reidentifies them when they play the game again on another day.
We follow the terminology used by the literature \citep{Rogers2015,Miller2022CombiningBiometrics,Miller2022TemporalBiometrics}, which describes the data collected during on-boarding of users as \textit{enrollment} and the data collected for later identification as \textit{use-time} data.
In our analyses, we take care to simulate a real-world user identification scenario:
the amount for both, enrollment and use-time data that can be captured per user can be very different depending on the individual use case.
Therefore, we limit the available amount of enrollment and use-time data to different lengths and vary this limit in our experiments (e.g., identifying a user within 5 minutes of use-time data based on 10 minutes of enrollment data).
More specifically, we evaluate enrollment and use-time data of 1, 5, 10, 15, 20, and 25 minutes per user, as well as all available data.
This provides us with evidence of how well the evaluated models perform in different use cases.
In the following, we describe different aspects of our experimental setup in more detail.

\subsubsection{Data Preprocessing}
\label{data-preprocessing}
We chose a sequence length of 300 frames with a frequency of 15 frames per second (fps) as input for the neural networks, which resembles a duration of $\frac{300\text{frames}}{15\text{fps}} = 20$ seconds. 
This is based on the following considerations.
First, we prefer shorter sequences, since the training time considerably increases with larger sequence lengths, especially for the GRU architecture.
Second, \cite{Rack2022ComparisonSequences} demonstrate that it is preferable for samples to cover a longer duration with a lower frequency than a shorter duration with a higher frequency. 
Lastly, 15 fps matches the frequency of the first half of the dataset (see \ref{sec:recording}).
Consequently, we resample the tracking data to a constant 15 fps for all users and then sample individual sequences of 300 frames as model inputs.

From each session, we remove the first and the last minute to remove potential artifacts that could have been caused by participants still getting ready or the recording stopping just after exiting the game.
Then, we divide the dataset into training, validation and testing sets.
In our case we consider session one for enrollment, hence we divide that session into training and validation split:
we select the last 5 minutes from session one for validation, and the rest for training.
The use-time data serves for the final evaluation, so we use entire session two for the test set.

\subsubsection{Evaluation Metric}

For evaluation, we consider the \textit{macro averaged accuracy}, which reports the ratio of correct predictions to the overall predictions made.
We use macro-averaging, which takes the accuracy of each class (i.e., user) individually and then computes the average score over all classes.
This way we account for eventual class imbalances, even though the footage for each user is fairly balanced.

During our analysis, we compare different lengths of enrollment data $t_{\text{enr}}$ per user with different lengths of use-time data $t_{\text{use}}$ per user.
To report the accuracy score for the various combinations, we use the format \acc{t_\text{enr}}{t_{\text{use}}}.
For example, if a model correctly identified 77\% of all 5 minute sequences from the test (i.e., use-time) data after being trained on 10 minutes of enrollment data per user, we write \acc{10}{5} $= 0.77$.

For the hyperparameter search, we employ the \textit{minimum accuracy} as key metric to compare models.
We do not use the macro averaged accuracy here, because we noticed during preliminary runs that models with similar validation accuracies can still be different in how consistently they identify each class.
Some models identified some classes very well, but scored very low for a few, while others performed well for all classes.
We think that the latter is preferable, since this allows to eventually identify each class reliably given enough use-time data.
The minimum accuracy considers this by taking the individual accuracies of each class and returning the score of the class with the lowest accuracy.
This way we prefer models that generalize well for all classes, and dismiss models that neglect individual classes.

\subsubsection{Hyperparameter Search}
\label{sec:hyperparameter_search}
We perform a hyperparameter search for the architectures in combination with each data encoding, which results in six separate searches (i.e., (GRU, CNN) $\times$ (BR, BRV, BRA)).
\Cref{tab:hyperparameters} lists the parameters we considered for both architectures.

We use the online service Weights and Biases \citep{wandb}.
The service monitors the training procedure and coordinates the search.
Here, we select the Bayesian search method, which observes a target metric during the search to systematically propose new configurations.
As target metric we use the minimum accuracy.

%% file: sections/04_results.tex
\subsection{Results}

We performed about 100 individual training runs for each architecture and data encoding combination, totaling 600 runs.
For each combination we pick the configuration that achieved the highest minimum accuracy score on the validation data.
The final configurations are listed in \Cref{tab:configurations}.

First, we compare the six models on the test data and pick the best model for further analysis.
As can be expected, identification accuracy improves when more use-time data is considered for a prediction.
This becomes visible in \Cref{fig:sequence_accuracies} which shows the correlation between amount of use-time data and identification accuracy for each model.
Overall, CNNs worked better than GRUs, and BRA worked better than BRV and BR.
The CNN+BRA combination performed best and achieved 95\% accuracy within 2 minutes (\acc{all}{2}) and 99\% accuracy within 7 minutes (\acc{all}{7}) of use-time data.
The CNN architecture not only outperformed the GRU architecture, but also trained much faster: the CNN+BRA model was trained within 2 hours, while the GRU+BRA model required 11.5 hours.
Since the BRA+CNN model worked best, we select it for further analysis.

Up to this point we used entire session one for enrollment, which is about 45 minutes of training data per user.
Depending on the target use case, it might not always be feasible to capture that much footage for each user.
Therefore, we re-train the model on shorter lengths of enrollment data, ranging from 1 to 25 minutes.
The starting point of the enrollment data within the recording of each user was selected randomly, so we repeat the training for each enrollment length 5 times and report the mean values in \Cref{fig:accuracies}.

Having a larger enrollment set appears to be more important than having a larger use-time set.
The accuracy for 1 minute enrollment and 10 minutes of use-time achieves only \acc{1}{10}$=10\%$, while 10 minutes enrollment and 1 minute of use-time achieves \acc{10}{1}$=54\%$. 
Doubling the amount of enrollment data from 1 to 2 minutes already improves the accuracy significantly to \acc{2}{10}$=29\%$.
This implies that if there is too little enrollment data, the model cannot make accurate predictions, even with more use-time data.
For instance, there is little difference between 1 minute and 25 minutes of use-time data, as long as there is only 1 minute of enrollment data (\acc{1}{1}=6\% and \acc{1}{25}=13\%).
However, after reaching a certain amount of enrollment data, lower prediction accuracies can be compensated with more use-time data, e.g., \acc{5}{1}$= 37\%$ compared to \acc{5}{25}$= 71\%$.

Random factors, such as the random initialization of the network and the GPU architecture, can impact the outcome of the training process.
To test its robustness against these factors, the CNN+BRA model was retrained 10 times with different seeds for random initialization.
The identification accuracy of individual input sequences, (i.e., \acc{all}{20s}), varied by only 1.7 percentage points, ranging from 76.6\% to 78.3\%. 
In comparison, a similar test during the investigation of the results of \cite{Liebers2021} resulted in variations of up to 15 percentage points.

%% file: sections/05_discussion.tex
\section{Discussion \& Conclusion}

With this article, we introduce a new dataset with behaviometric and biometric data from VR users to address a critical issue in this field.
The lack of such datasets has hindered the development of larger and more accurate machine learning models, making it difficult to compare new methods to current ones or reproduce previous solutions.
We believe that publishing our dataset is an important step towards improving the reproducibility and reliability of machine learning methods in XR user identification and authentication research.

Both tested state of the art deep learning models were able to identify users in session two after training on session one.
All models performed much better than a random classifier would which validates that the tracking data provides identifying user-specific characteristics.
The best model was a CNN architecture that could identify 95\% of all 2 minute sequences from session two correctly when trained on all data from session one.
The CNN architecture also was trained fastest and generally only needed about half the time to reach peak performance compared to the RNN.
This performance deviates just slightly (i.e., below $\pm 1$ percentage point of accuracy) if the model is trained with different seeds, which indicates that models can be trained and evaluated in a robust way with the used dataset and methods.

The results also reveal that the acceleration encoding (BRA) yields superior performance compared to BR and BRV.
This finding supports the hypothesis put forth by \cite{Rack2022ComparisonSequences} that abstracting non-motion-related information can enhance the generalization capabilities of the model.
We hypothesize that more sophisticated encoding techniques could potentially lead to even better identification models.
Thus, this remains a promising area for future research.

It is crucial for the scientific community to have access to diverse and well-documented datasets in order to facilitate progress and advance research.
Therefore, we strongly urge authors of future work to make their datasets publicly available, along with any relevant information about data collection and processing.
This will enable other researchers to reproduce the results, conduct further analysis, and build on previous work, ultimately leading to a deeper understanding of the field.

A limitation of the dataset is that it contains only marginally contextualized motion sequences of users being passively watched while they engage with the VR application.
While this is ideal to explore identification models that can work by passively watching XR users, it is less suited to explore active verification use cases, as these usually expect the user to actively provide a short input sample.
However, \cite{Rack2023} show that modern embedding-based models are able to generalize once trained on the ``Who Is Alyx?'' dataset to a different dataset exhibiting different motions.
Consequently, one or more large datasets, like ``Who Is Alyx?'', could potentially be used to train powerful multi-purpose deep learning models, which can then be used not only for identification, but also for verification of users.

This article introduces the ``Who Is Alyx?" dataset, specifically designed for investigating user identification based on motion in XR environments.
Using two state-of-the-art deep learning architectures, we have established an initial benchmark that future research can build upon.
The breadth of our dataset extends beyond the scope of this article, as ``Who Is Alyx?'' encompasses a wide array of unexplored features that we believe hold promise for further exploration.
We invite the scientific community to utilize this dataset in their research and welcome inquiries, feedback, and suggestions from researchers working with our dataset.
Furthermore, we have plans to continue extending ``Who Is Alyx?'' and collect data from other scenarios as well.
In doing so, we hope to facilitate the creation of additional datasets encompassing a wider range of XR contexts, pushing towards novel and reliable user recognition systems to improve security and usability in XR.






%% file: sections/99_figures_and_tables.tex
\clearpage
\section*{Tables}

\begin{table}[H]
    \caption{Publicly available XR motion datasets.}
    \label{tab:existing_datasets}
    \centering
    \begin{tabular}{l|lllll}
    \toprule
        \textbf{Dataset}      & \cite{Liebers2021} & \cite{Rack2022ComparisonSequences} & \cite{Miller2022CombiningBiometrics} & \cite{BOXRR-23}            & ours \\ \midrule
        \textbf{Users}        & 16                 & 34                                 & 41 and 16                            & 105,852                    & 71 \\ \midrule
        \textbf{Footage/user} & 8 min.             & 60 min.                            & 2 min.                               & 36 min.* & 90 min. \\ \midrule
        \textbf{Sessions}     & 2                  & 1                                  & 2                                    & unknown                          & 2 \\ \midrule
        \multicolumn{6}{r}{\tiny *median value} \\ 
    \end{tabular}

\end{table}

\begin{table}[H]
\caption{Overview of recorded data.}
\label{tab:recording-overview}
\begin{tabular*}{\linewidth}{@{}llrr@{}}
\toprule
\textbf{Device}                       & \textbf{Sampling Rate} & \textbf{1 Session} & \textbf{2 Sessions} \\ \midrule
\textbf{HTC Vive Pro (motions)} &    combined               & 5                  & 71                  \\
                                      & 15 Hz            & 4                  & 37                  \\
                                      & 90 Hz                  & 1                  & 34                  \\ \midrule
\textbf{Eye-tracking}                 &     80 Hz                   & 8                  & 67                  \\ \midrule
\textbf{Empatica E4 (wristband)}      &                        & 8                  & 27                  \\
– Acceleration                        & 32 Hz                  &                    &                     \\
– Heart Rate                          & 1 Hz                 &                    &                     \\
– Electrodermal Activity              & 4 Hz                   &                    &                     \\
– Photoplethysmography                & 64 Hz                  &                    &                     \\
– Inter-beat interval                 & 64 Hz                  &                    &                     \\
– Peripheral Body Temp.         & 4 Hz                   &                    &                     \\ \midrule
\textbf{Polar H10 (chest strap)}      &                        & 3                  & 31                  \\
– Acceleration                        & 200 Hz                 &                    &                     \\
– Electrocardiogram                   & 130 Hz                 &                    &                     \\ \bottomrule
\end{tabular*}
\end{table}

\begin{table}[H]
\caption{Hyperparameter search spaces.}
\label{tab:hyperparameters}
\begin{tabularx}{\linewidth}{p{1.5cm}ll}
\toprule
                         & \textbf{hyperparameter} & \textbf{search space}  \\ 
\midrule
\textbf{GRU}             & hidden size             & [20, 450] neurons \\
                         & number of layers        & [1, 8]           \\
                         & dropout                 & [0, 0.6] \\
                         & learning rate           & [0.0001, 0.01] \\
\midrule
\textbf{CNN}             & kernel sizes            & [2, 9] \\
                         & number of layers        & [1, 8] \\
                         & channel sizes           & [10, 500] \\
                         & dropout                 & [0, 0.6] \\
                         & learning rate           & [0.0001, 0.01] \\
\bottomrule
\end{tabularx}
\end{table}

\begin{table}[H]
\tiny
\caption{Final model configurations after hyperparameter search.}
\label{tab:configurations}
\input{results/generated/combined_configurations}
\end{table}

\clearpage
\section*{Figures}

\setcounter{figure}{2}
\setcounter{subfigure}{0}
\begin{subfigure}[h!]
\setcounter{figure}{2}
\setcounter{subfigure}{0}
  \begin{minipage}[b]{.3\textwidth}
  \includegraphics[width=\textwidth]{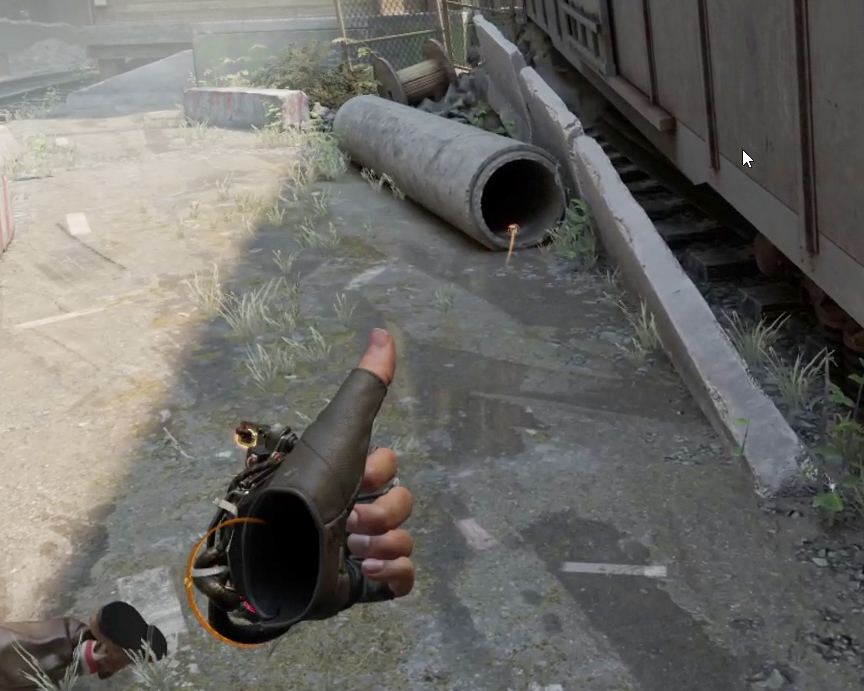}
  \caption{Gravity grab: player targets object.}
  \label{fig:gravity_grab_1}
  \end{minipage}%
\setcounter{figure}{2}
\setcounter{subfigure}{1}
  \begin{minipage}[b]{.3\textwidth}
  \includegraphics[width=\textwidth]{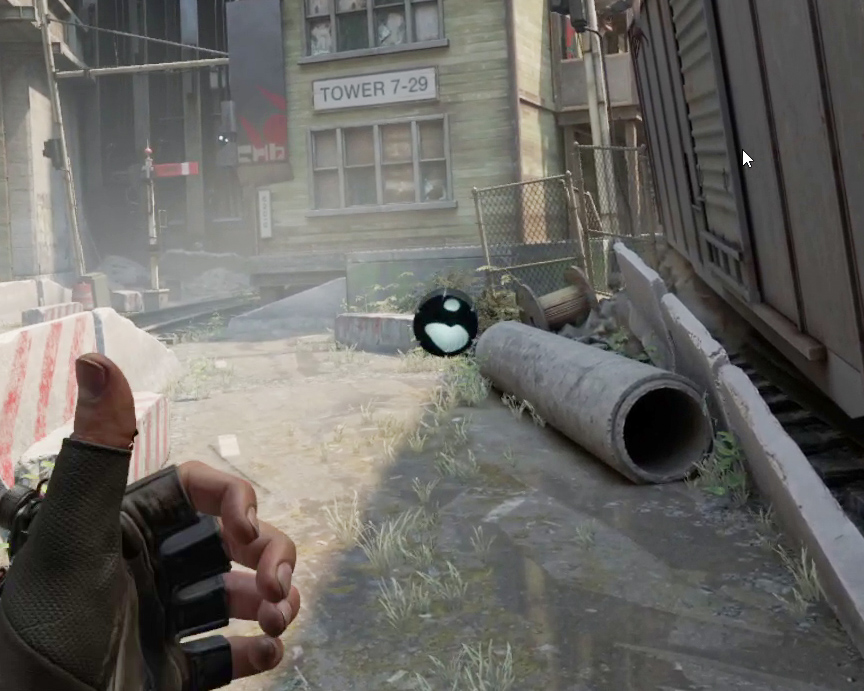}
  \caption{Gravity grab: hand flip propels object.}
  \label{fig:gravity_grab_2}
  \end{minipage}%
\setcounter{figure}{2}
\setcounter{subfigure}{2}
  \begin{minipage}[b]{.3\textwidth}
  \includegraphics[width=\textwidth]{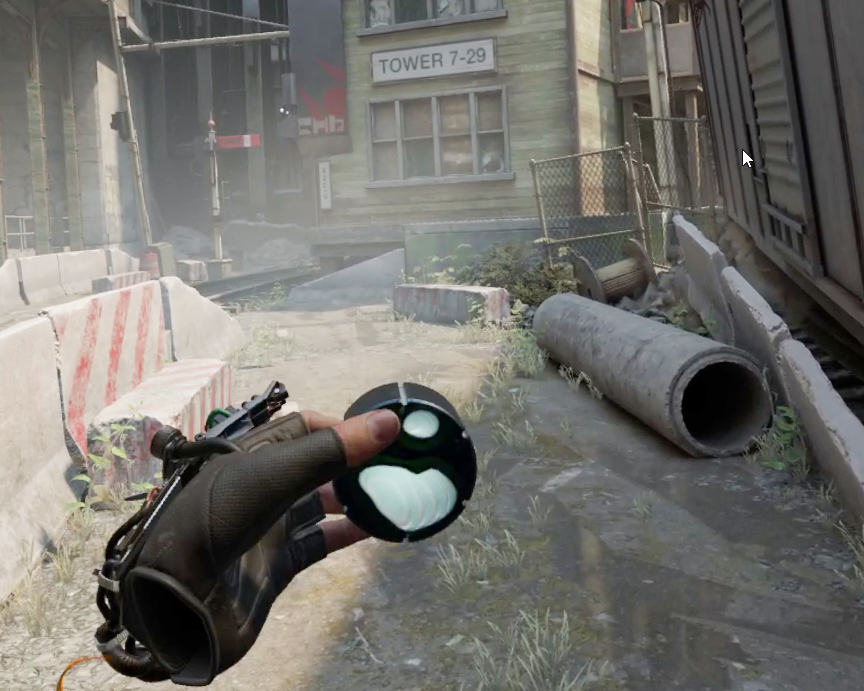}
  \caption{Gravity grab: player catches object.}
  \label{fig:gravity_grab_3}
  \end{minipage}%
\setcounter{figure}{2}
\setcounter{subfigure}{3}
  \begin{minipage}[b]{.3\textwidth}
  \includegraphics[width=\textwidth]{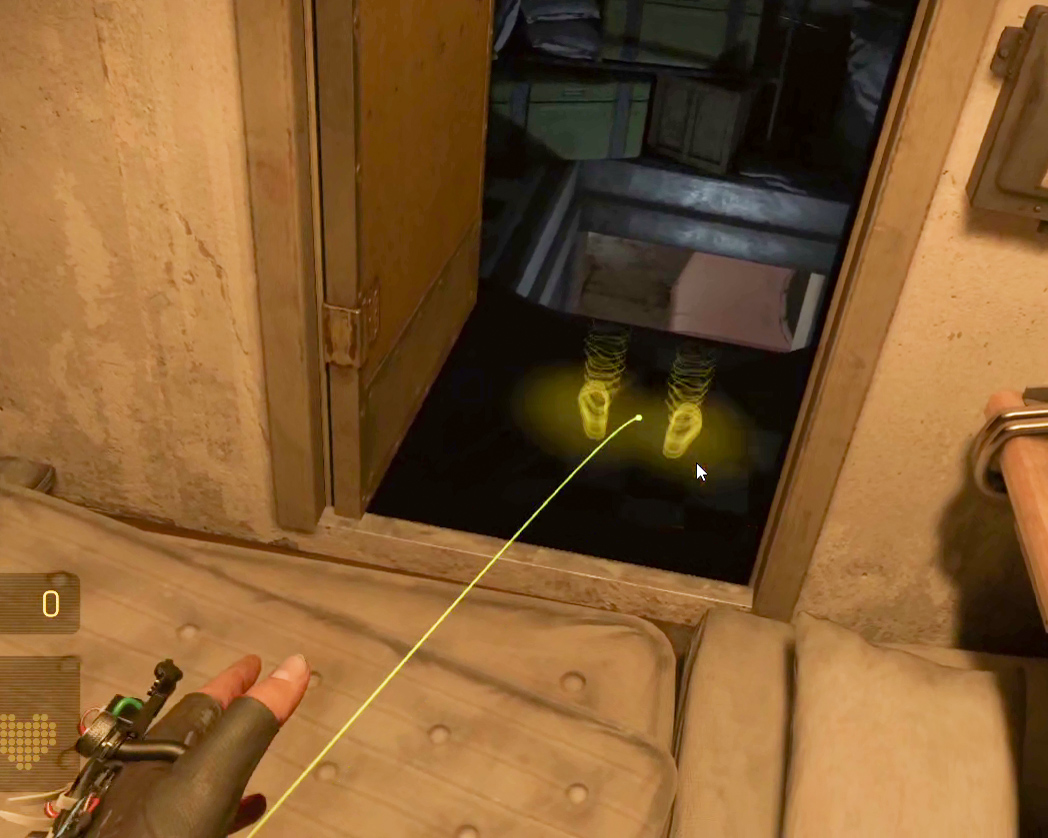}
  \caption{Player teleports by pressing the trackpad of the left controller.}
  \label{fig:teleportation}
  \end{minipage}%
\setcounter{figure}{2}
\setcounter{subfigure}{4}
  \begin{minipage}[b]{.3\textwidth}
  \includegraphics[width=\textwidth]{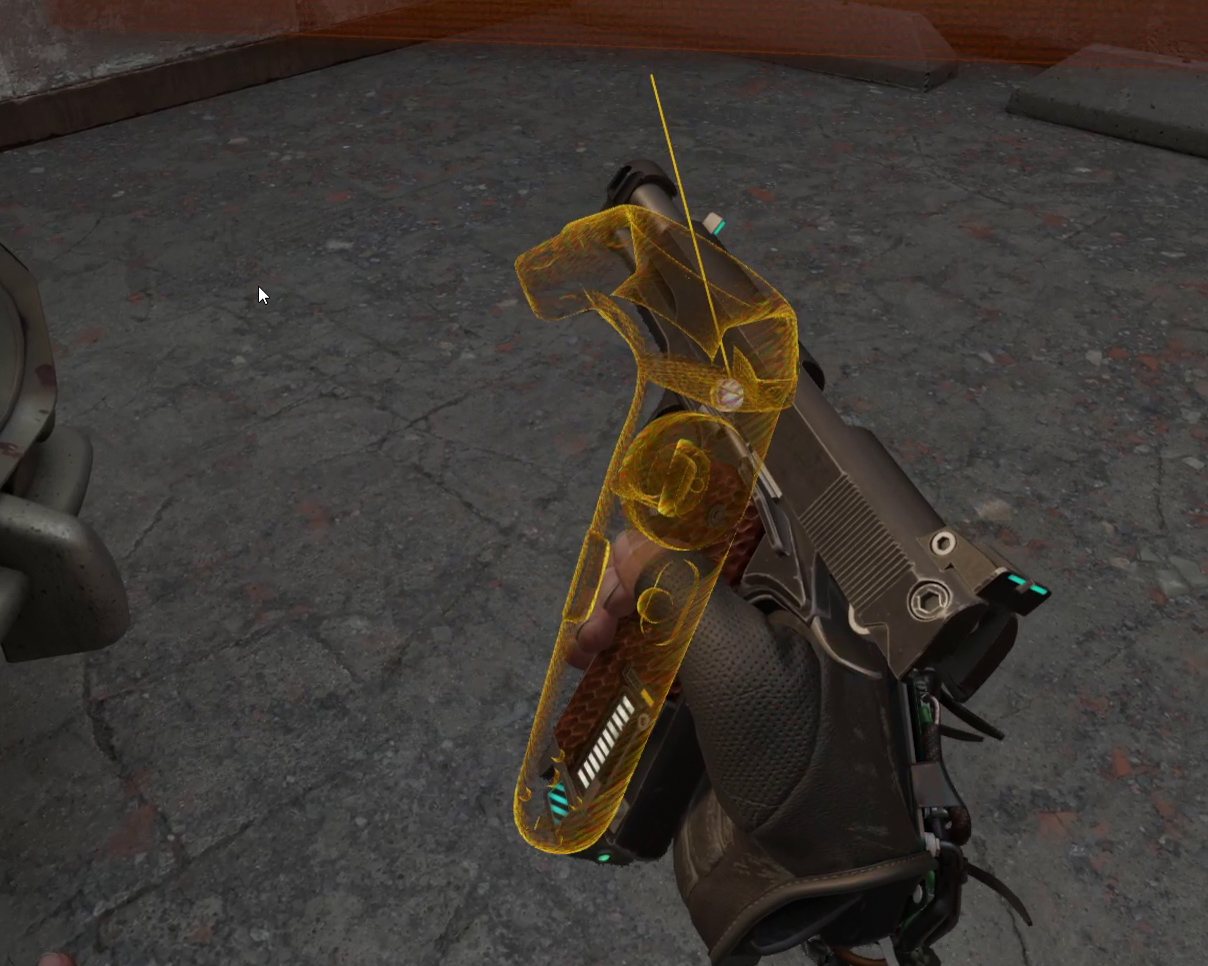}
  \caption{Right hand holds weapons and tools.}
  \label{fig:weapon}
  \end{minipage}%
\setcounter{figure}{2}
\setcounter{subfigure}{5}
  \begin{minipage}[b]{.3\textwidth}
  \includegraphics[width=\textwidth]{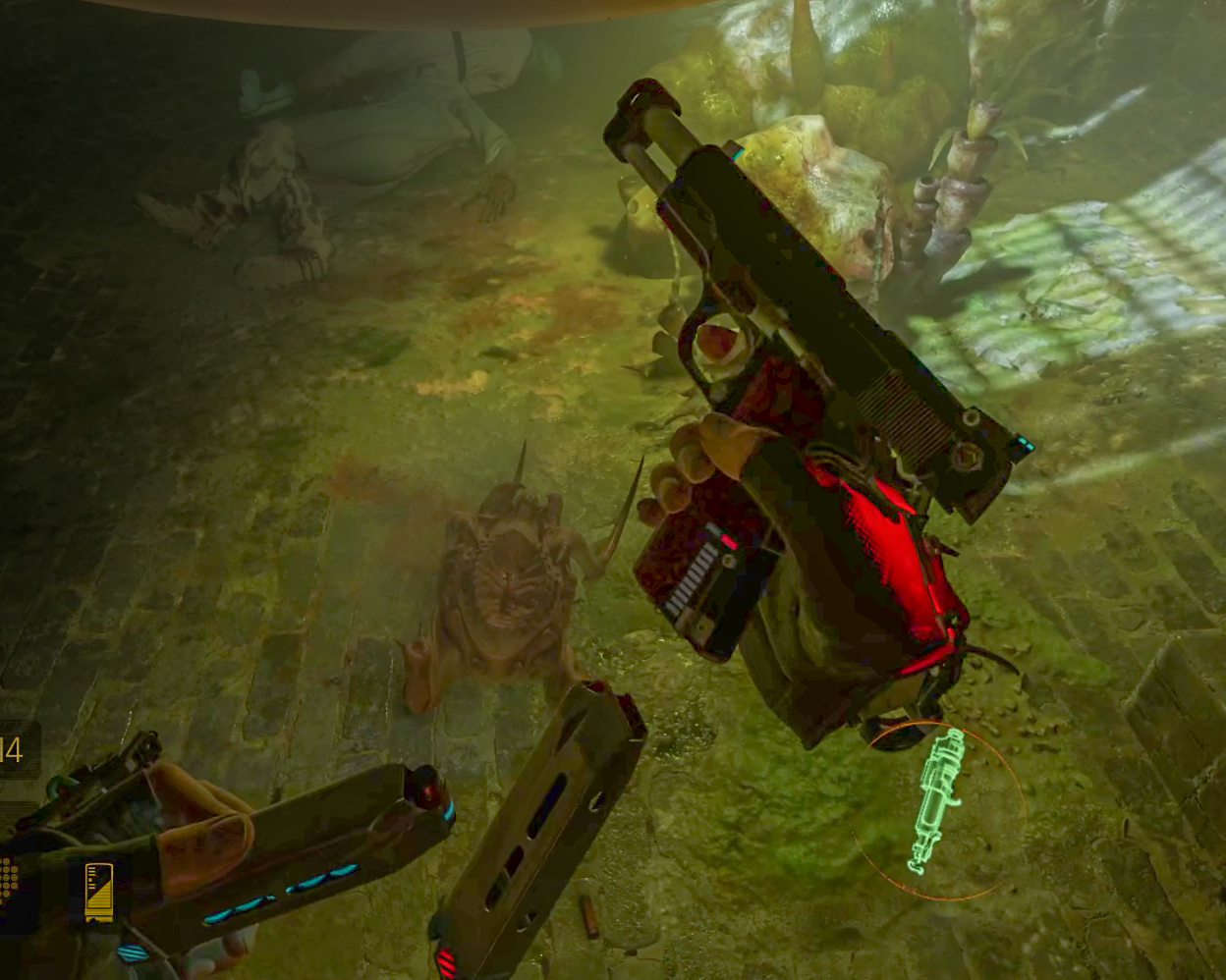}
  \caption{Player drops old mag to replace it with new one to reload weapon.}
  \label{fig:weapon_reload}
  \end{minipage}%
\setcounter{figure}{2}
\setcounter{subfigure}{6}
  \begin{minipage}[b]{.3\textwidth}
  \includegraphics[width=\textwidth]{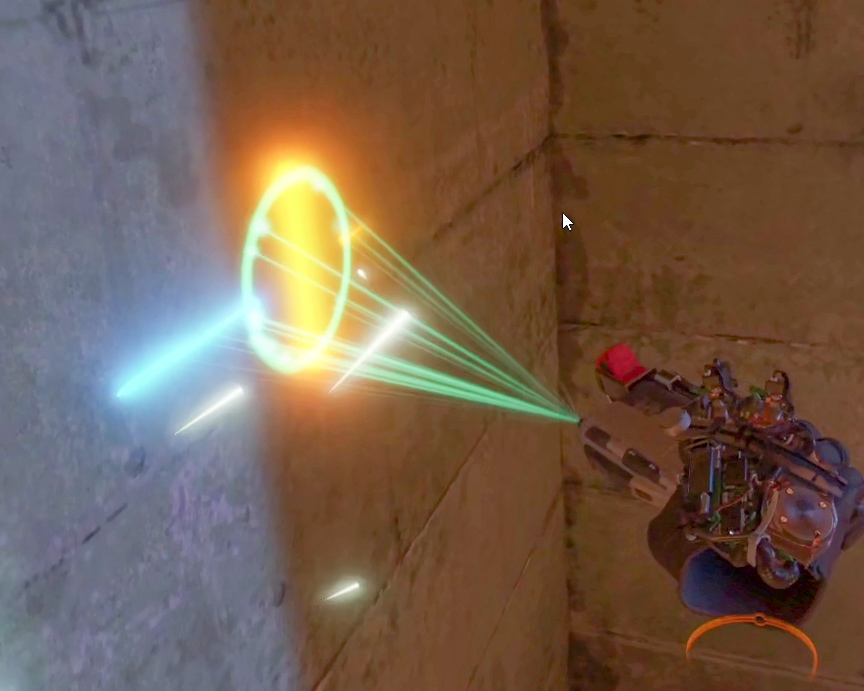}
  \caption{Player has to find and fix circuits in the wall to unlock barriers.}
  \label{fig:circuit}
  \end{minipage}%
\setcounter{figure}{2}
\setcounter{subfigure}{7}
  \begin{minipage}[b]{.3\textwidth}
  \includegraphics[width=\textwidth]{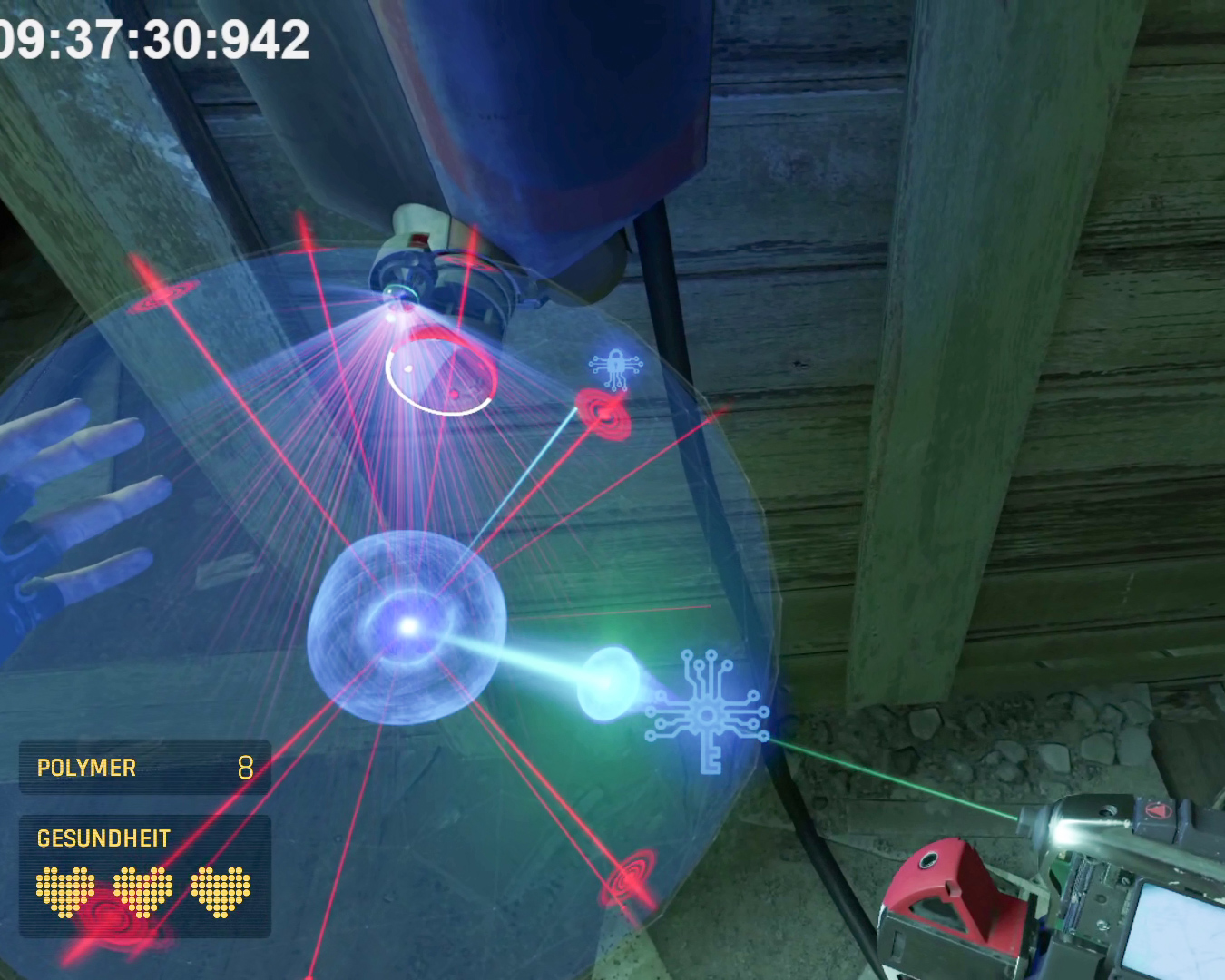}
  \caption{3D puzzles like this involve both hands and unlock bonuses.}
  \label{fig:puzzle}
  \end{minipage}%
\setcounter{figure}{2}
\setcounter{subfigure}{8}
  \begin{minipage}[b]{.3\textwidth}
  \includegraphics[width=\textwidth]{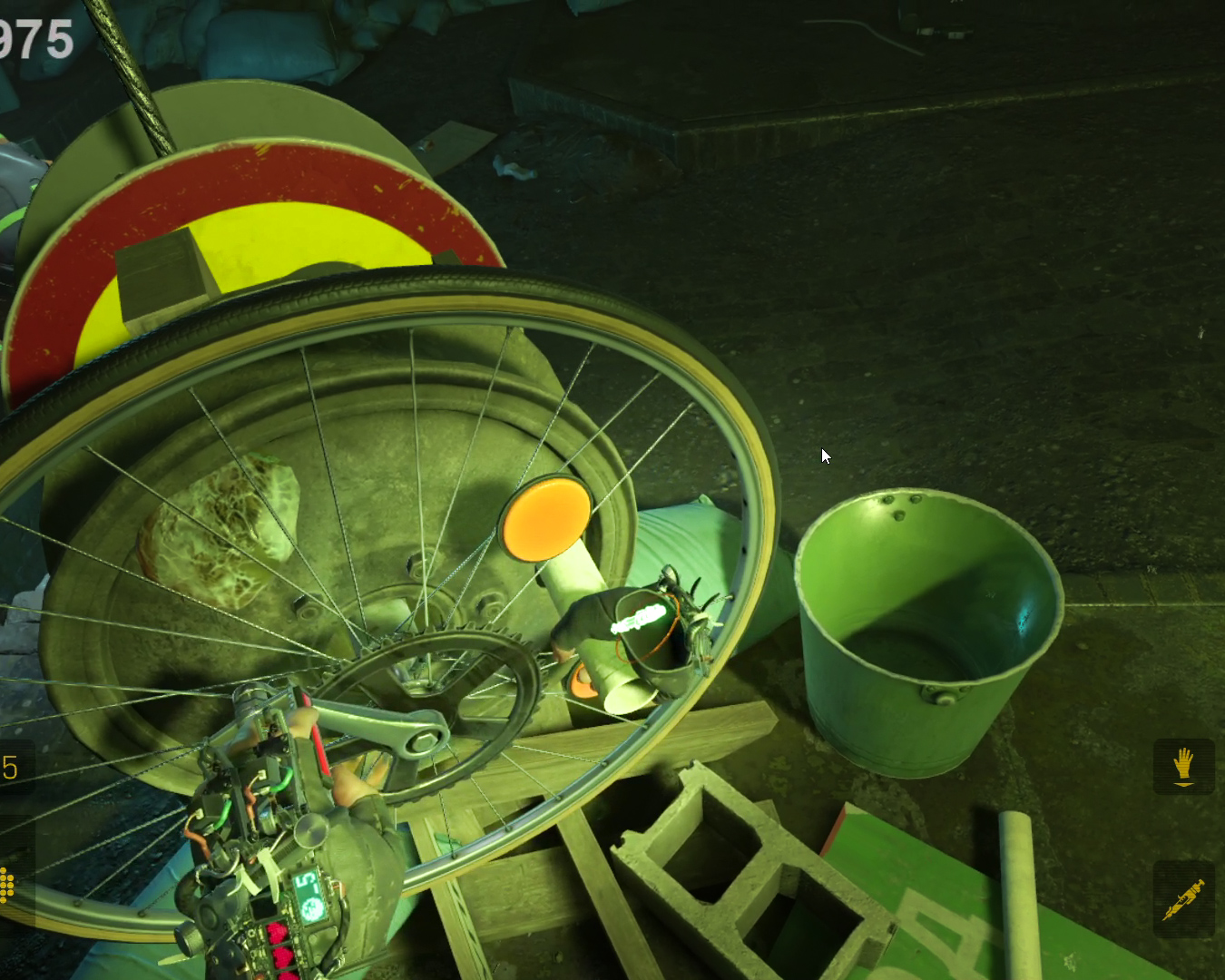}
  \caption{Player rolls up a rope and uses a tube to prevent it from unrolling again.}
  \label{fig:unique_puzzle}
  \end{minipage}%
  \setcounter{figure}{2}
  \setcounter{subfigure}{-1}
  \caption{Different scenes from the game, illustrating the wide range of different actions required from the players.}
  \label{fig:teaser}
\end{subfigure}%

\begin{figure}[H]
    \centering
    
    \input{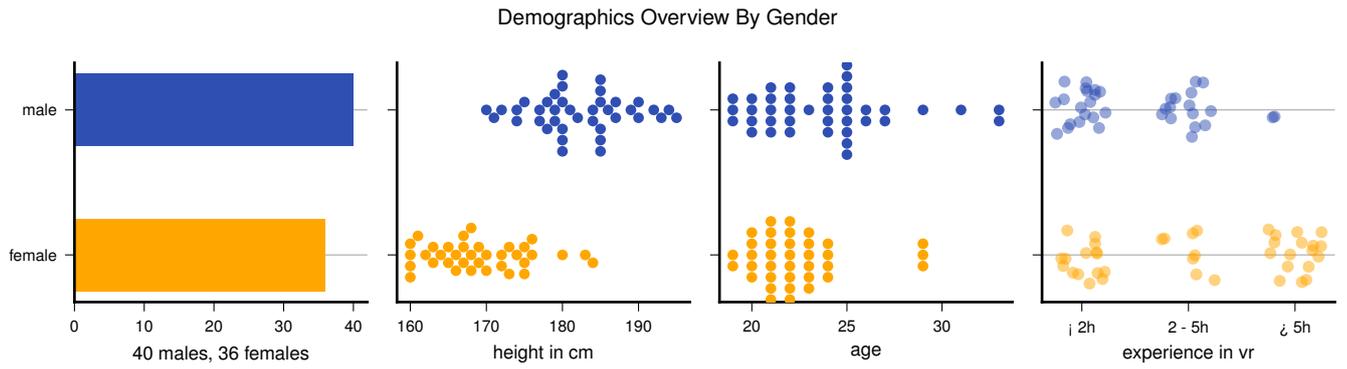}
    \caption{Overview of collected demographic data, divided by gender.}
    \label{fig:demographics}
\end{figure}

\begin{figure}[H]
    \centering
    
    \input{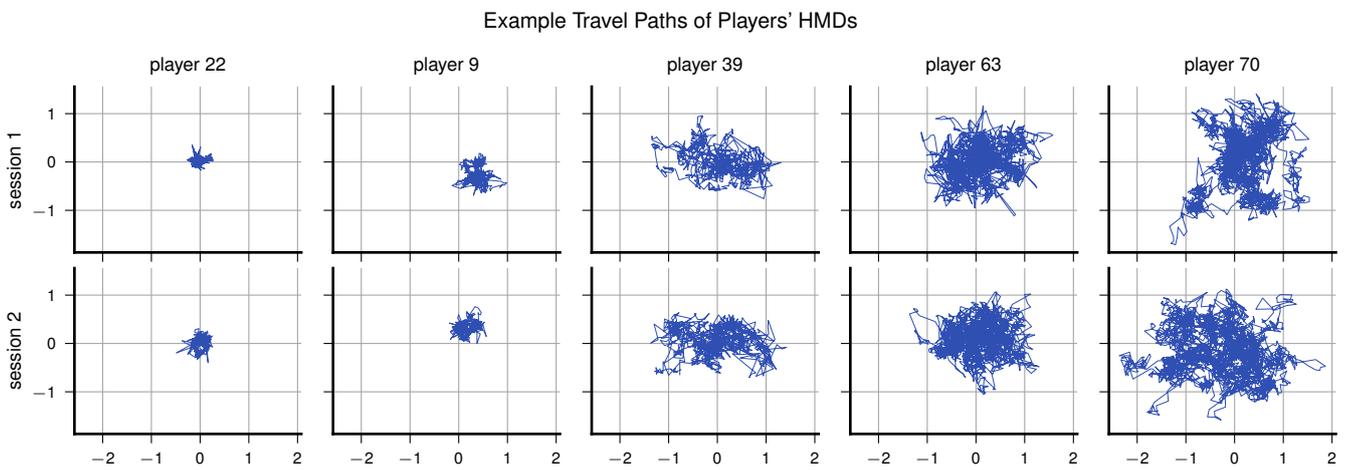}
    \caption{Lateral travel paths of five players, ordered by avg. meters traveled per minute; unit of axes is `meters'.}
    \label{fig:player_travel_path}
\end{figure}

\begin{figure}[H]
    \centering
    \input{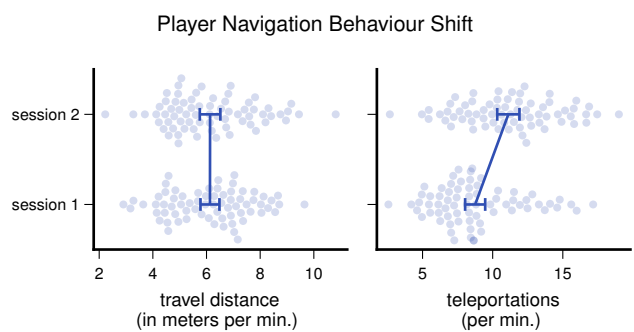}
    \caption{Differences in navigation behaviour; dots denote players, error bars mark 95\% confidence of the means.}
    \label{fig:navigation_shift_swarmplot}
\end{figure}

\begin{figure}[H]
    \centering
    \input{results/generated/model_accuracy.pgf}
    \caption{CNN+BRA accuracy for different enrollment and use-time data.}
    \label{fig:accuracies}
\end{figure}

\begin{figure}[H]
    \centering
    \input{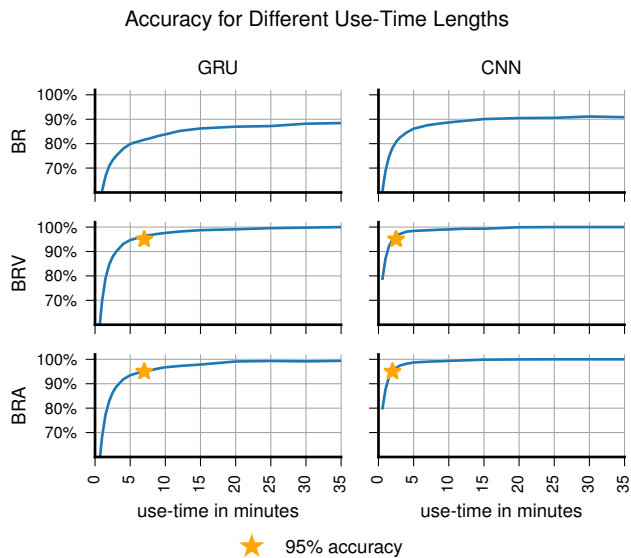}
    \caption{Model performances when trained on all enrollment data (i.e., entire session one).}
    \label{fig:sequence_accuracies}
\end{figure}

%% file: results/generated/combined_configurations.tex
\centering
\begin{tabular}{l|rrrr|rrrrr}
\toprule
 & \multicolumn{4}{c}{\textbf{GRU}} & \multicolumn{5}{c}{\textbf{CNN}} \\
 & learning rate & layers & hidden size & dropout & learning rate & layers & kernel size & channel sizes & dropout \\
\midrule
\textbf{BR} & 0.0005 & 3 & 300 & 0.32 & 0.002 & 4 & 3 & 400,580,841,1219 & 0.46 \\
\textbf{BRV} & 0.0007 & 4 & 250 & 0.32 & 0.002 & 3 & 3 & 400;480;576 & 0.42 \\
\textbf{BRA} & 0.0005 & 4 & 400 & 0.10 & 0.002 & 4 & 3 & 600;900;1350;2025 & 0.44 \\
\bottomrule
\end{tabular}